\documentclass[showpacs,amsmath,amssymb, prl,twocolumn]{revtex4}
\usepackage{graphicx}
\usepackage{dcolumn}
\usepackage{appendix}
\usepackage{bm}

\begin{document}

\title{Noise-assisted Thouless pump in elastically deformable molecular junctions}

\author{C.A. Perroni$^{1}$, F. Romeo$^{3}$, A. Nocera$^{2}$, V. Marigliano Ramaglia$^{1}$, R. Citro$^{3}$, and V. Cataudella$^{1}$ }

\affiliation{$^{1}$ CNR-SPIN and Universita' degli Studi di Napoli ''Federico II''\\
Complesso Universitario Monte S. Angelo, Via Cintia, I-80126
Napoli, Italy \\
$^{2}$ Department of Physics, Northeastern University, Boston, Massachusetts 02115, USA \\
$^{3}$ Dipartimento di Fisica "E. R. Caianiello"and CNR-SPIN, Universita' degli Studi di Salerno \\
Via Giovanni Paolo II, I-84084
Fisciano, Italy}

\begin {abstract}
We study a Thouless pump realized with an elastically \textit{deformable quantum dot} whose center of mass follows a non-linear stochastic dynamics. The interplay of noise, non-linear effects, dissipation and interaction with an external time-dependent driving on the pumped charge is fully analyzed. The results show that the quantum pumping mechanism not only is not destroyed by the force fluctuations, but it becomes stronger when the forcing signal frequency is tuned close to the resonance of the vibrational mode. The robustness of the quantum pump with temperature is also investigated and an exponential decay of the pumped charge is found when the coupling to the vibrational mode is present. Implications of our results for nano-electromechanical systems are also discussed.
\end {abstract}

\maketitle



\section{Introduction}
Modern nanoelectronic technologies open the possibility to test several transport regimes and quantum protocols at the boundary of quantum and classical realm.
In this intermediate regime the reduced device dimensions (smaller than the electronic mean free path) allow to preserve the quantum coherence of the particles flux evidencing an emergent mesoscopic behavior. In mesoscopic systems manmade nanocircuits can interact coherently with organic molecules giving origin to hybrid structures. A sub-class of these organic-inorganic hybrids is represented by the hetero-elastic nanodevices in which an elastic response is observed\cite{di_ventra}. In these systems the charge density interacts with internal degrees of freedom that determine the elastic response of the device giving rise to new functionalities \cite{Clerk}. The simplest system belonging to this class of nanodevices is realized by an organic molecule (elastically soft part) connected to two external inorganic leads (elastically hard part). A minimal description for the above system is given in terms of the center of mass dynamics of the soft part (considered as classical) and its coupling to the charge state activated during the transport (requiring a quantum description)\cite{Gorelik,Gorelik13,Armour}.\\

An interesting situation is the one of a Thouless pump\cite{thouless83} realized by using an hetero-elastic system.
Experimentally, organic materials like carbon nanotubes\cite{leek05,buitelaar08}, and graphene nanoribbons\cite{chen,low12} have been recently employed in the realization of pumping nanodevices exploiting their mechanical properties.
In conventional quantum pumping (\textit{\`{a} la} Thouless) the adiabatic time-modulation of two external parameters $X_{1,2}(t)=X^{0}_{1,2}+\delta X_{1,2} \sin(\omega t+\phi_{1,2})$ parametrically affects the system Hamiltonian $\mathcal{H}(X_1(t),X_2(t))$ and produces a dc particle current proportional to the quantity $\omega \delta X_1 \delta X_2 \sin(\varphi)$, $\omega$ being the pumping frequency and $\varphi=\phi_1-\phi_2$ the phase difference between the driving signals \cite{brouwer98,switkes}.
This current-phase relation (CPR) is substantially modified when the scattering region reacts to the modifications of the internal charge state by adjusting its spatial configuration \cite{Oppen}. The latter situation has been studied in Ref.\cite{deformable_qd} where an elastically \textit{deformable quantum dot} (DQD) has been studied. Assuming a linear \textit{deterministic} dynamics for the mechanical deformation of the scattering region (harmonic oscillator paradigm),  it has been demonstrated that the CPR is modified by the presence of a dynamical phase shift $\phi_D$ of the response function of the harmonic oscillator. Furthermore, when the DQD model is generalized to include a weak non-linearity in the oscillator dynamics \cite{memory_qd} interesting memory effects take place. In general, as also observed in Ref.\cite{parasitic_int_qd}, the pumping current is feeded by all the phase differences characterizing the scattering region and thus the presence of an hidden classical dynamics coupled to the transport determines a strong deviation from the conventional CPR of the Thouless pumping.\\


In this work, we study a Thouless pump realized with an elastically DQD whose dynamics is affected by  stochastic forces induced by quantum and thermal fluctuations due to the electron charging of the quantum dot. Exploiting a non-equilibrium Green function approach that leads to a Langevin dynamics for the elastic part of the system, as formulated in Ref. \cite{alberto,alberto0,alberto1,alberto2}, the pump response has been fully analyzed. The attention has been mainly focused on how the pumped charge depends on temperature and on the driving frequency. We show that, although pumping is strongly reduced with temperature, as expected, close to the resonance conditions (external driving frequency close to the resonator's frequency) the pumping can be enhanced of more than one order of magnitude leading to measurable pumping effect up to very high temperatures ($K_B T$ of the order or larger than the hybridization energy between the leads and the nanotube). Moreover we find that the temperature decay of the pumped charge in the presence of the coupling to a vibrational mode is slower than that obtained in the non-deformable Thouless pump. It is also interesting to note that, differently from the expectation, the resonance takes place at a strongly renormalized value of the bare resonator frequency despite the moderate electron-phonon coupling. Finally, we observe  non linear features in the resonator dynamics close to the resonance.

The paper is organized as follows: in Sec. II the Anderson-Holstein model is introduced; in Sec. III the adiabatic regime for quantum pumping is considered in the framework of a non-equilibrium Green functions approach and the Langevin dynamics is derived; in Sec. IV we discuss the results for the pumped charge and then give the conclusions. There are two appendices: Appendix 1 on the numerical procedure and Appendix 2 on the zero temperature behavior of the system.

\begin{figure}
\centering
\includegraphics[scale=0.55]{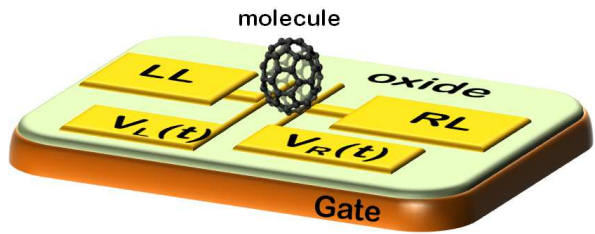}
\caption{(Color online) Scheme of the device studied in this work. The Left Lead (LL) and the Right Lead (RL) are kept at the same chemical potential. The pumping signals are applied using the gates $V_{L/R}(t)$, while the back gate (Gate) induces a $V_G$ shift to the molecular energy level. }\label{pisto}
\end{figure}

\section{The Anderson-Holstein (AH) model}
The spinless Anderson-Holstein model is the simplest model of a
molecular junction including the effect of electron-oscillator
interaction (see Fig. 1 for a scheme of the device). The molecule is modeled as an electronic level
interacting locally with a single vibrational mode and whose Hamiltonian $\hat{H}$ is given by:
\begin{equation}
\hat{H}={\hat H}_{el}+ {\hat H}_{osc} + {\hat H}_{int},\label{Htot}
\end{equation}
where the electronic system is described by the standard junction Hamiltonian
$\hat{H}_{el}$:
\begin{equation}
\hat{H}_{el}=V_G {\hat d^{\dag}}{\hat d}
+ \sum_{q,\alpha} \left[ V_{q,\alpha}(t){\hat c^{\dag}_{q,\alpha}}{\hat d}+ h.c. \right] +
\sum_{q,\alpha}\varepsilon_{q,\alpha}{\hat c^{\dag}_{q,\alpha}}{\hat c_{q,\alpha}}.
\end{equation}
The molecular electronic level has energy $V_G$, determined by the gate voltage, and  ${\hat d^{\dag}} ({\hat d})$ are creation
(annihilation) operators. The
operators ${\hat c^{\dag}_{q,\alpha}} ({\hat c}_{q,\alpha})$
create (annihilate) electrons with momentum $q$ and energy
$\varepsilon_{q,\alpha}=\xi_{q,\alpha}-\mu_{\alpha}$ in the left
($\alpha=L$) or right ($\alpha=R$) free metallic leads. The
chemical potentials in the leads, $\mu_{L}$ and $\mu_{R}$, are
assumed to be equal: $\mu_{L}=\mu_{R}=0$ (absence of external voltage bias).
The leads will be considered as thermostats in equilibrium at temperature $T$.
The electronic tunneling between the molecular dot and a state $k$ in the lead $\alpha$
has the time dependent amplitude $V_{q,\alpha}(t)$.  For
the sake of simplicity, we will suppose that the density of states $\rho_{q,\alpha}$ for
the leads is flat within the wide-band approximation: $ \rho_{q,\alpha} \mapsto \rho_{\alpha}$,
$V_{q,\alpha}(t) \mapsto V_{\alpha} u_{\alpha} (t) $, with $u_{\alpha} (t)$ a periodic
function of time governing the strength of the pumping (see Fig. 1). Therefore, the time dependent
full hybridization width of the molecular orbital is
$\hbar \Gamma(t,t')=\sum_{\alpha } \hbar \Gamma_{\alpha} (t,t')=\sum_{\alpha } \hbar \Gamma_{\alpha} u_{\alpha}(t) u_{\alpha}(t')$, with $\hbar$ Planck constant and the
tunneling rate $\Gamma_{\alpha}=2\pi\rho_{\alpha}|V_{\alpha}|^{2}/\hbar$. In the following, we consider
the symmetric configuration: $\Gamma_L=\Gamma_R=\Gamma_0$.

In Eq.({\ref{Htot}), the Hamiltonian of the oscillator (or vibrational mode) is given by
\begin{equation}
{\hat H}_{osc}={{\hat p}^{2}\over 2m} + {1\over 2} k {\hat x}^{2},\label{Hosc}
\end{equation}
with ${\hat p}$ and ${\hat x}$ momentum and position operator, respectively, while $m$ is the effective mass, $k$ the spring constant, and
the oscillator frequency is $\omega_{0}=\sqrt{k/m}$.

In Eq.({\ref{Htot}), the interaction term ${\hat H}_{int}$ (typically of electrostatic origin \cite{alberto2}) is
provided by a simple linear coupling between the electron
occupation on the molecule, ${\hat d^{\dag}}{\hat d}$, and the
displacement $\hat{x}$ of the oscillator:
\begin{equation}
{\hat H}_{int}=\lambda {\hat x} {\hat n},\label{Hint}
\end{equation}
where $\lambda$ is the electron-oscillator coupling
strength and ${\hat n}={\hat d^{\dag}}{\hat d}$ is the density operator. In the following, the coupling between the electron system and the
vibrational mode will be often described in terms of the
electron-oscillator coupling energy $E_{ep}=\lambda^2/(2 k)$.

\section{Adiabatic regime}
In the following we will study the system under generic pumping strength and arbitrary electron-oscillator coupling
in the experimentally relevant limit $\hbar \omega_0 \ll K_B T$, with $K_B$ Boltzmann constant.
In this regime, the dynamics of the oscillator is classical.
Therefore, the electronic dynamics is equivalent to a time dependent single level
problem with energy $E_{0}(t)=V_G+\lambda x(t)$.
Using the Keldysh formalism\cite{Haug,Arrachea,Perroni} we can
solve the Dyson and Keldysh equations for the molecular Green
functions. Actually, the retarded molecular Green's function can be easily obtained as
\begin{equation}
G_{r}(t,t')=-{i \over\hbar} \theta(t-t')
e^{- \frac{i}{\hbar} \int_{t'}^{t} dt_{1} \left[ E_{0}(t_{1}) - i \hbar \Gamma (t_1) /2 \right] },\label{grtimedep}
\end{equation}
with $\theta(t)$ Heaviside function, and $\Gamma(t)=\Gamma(t,t)$. The lesser Green function can be evaluated exactly, in particular
that at equal time
\begin{equation}
G_{<}(t,t)=i\sum_{\alpha} \Gamma_{\alpha} \int_{- \infty}^{+ \infty}
{d (\hbar \omega) \over2\pi} f(\omega) |B_{\alpha}(\omega,t)|^2, \label{gless}
\end{equation}
with $f(\omega)$ Fermi function for both the leads $\alpha=R,L$, and the function
$B_{\alpha}(\omega,t)$ defined as
\begin{equation}
B_{\alpha}(\omega,t)= \int_{- \infty}^{+ \infty} dt_{1} u_{\alpha} (t_1) G_{r}(t,t_1)
e^{ i \omega (t-t_1)}.\label{balfa}
\end{equation}
It can be interpreted as the retarded Green function dressed with the pumping term $u_{\alpha}(t)$.

We analyze the electron system in the adiabatic regime for the electron dynamics, that is in the limit of slow
temporal perturbations:
$\omega_0 \ll \Gamma_0$ and $d \Gamma /dt \ll \Gamma_0^2$. Within this regime, one uses the following expansion for the pumping parameter:
$u_{\alpha}(t_1) \simeq u_{\alpha}(t)+ \dot{u}_{\alpha}(t)(t_1-t)$, with the dot indicating the time derivative.
Therefore, one can calculate the adiabatic expansion of the Green's function considering the explicit dependence of the electronic
quantities on the oscillator variables $x$,$v$, and its intrinsic dependence on time $t$ governed by the pumping terms $u_{\alpha}(t)$.

Within the adiabatic approximation one can derive the following expansion of the dot occupation $N(x,v,t)$:
\begin{equation}
N(x,v,t) =\langle \hat n\rangle(t) \simeq N^{(0)}(x,t)+N^{(1)}(x,v,t),
\end{equation}
where the zero order "static" term $N_{el}^{(0)}(x,t)$ is
\begin{equation}
N^{(0)}(x,t)= \int_{- \infty}^{+ \infty}
{d (\hbar \omega) \over 2\pi} f(\omega) \frac{\hbar  \Gamma(t)}{ ( \hbar \omega -\lambda x )^2+\frac{ [\hbar \Gamma(t)]^2 }{4}},\label{enne0}
\end{equation}
and the first order "dynamic" term $N_{el}^{(1)}(x,v,t)$ is
\begin{eqnarray}
N^{(1)}(x,v,t) =
\frac{\hbar}{2} \left[ \lambda v [\hbar \Gamma(t)] R(x,t) +  [\hbar \dot{\Gamma}(t)] R_1(x,t) \right] ,\label{enne1}
\end{eqnarray}
with
\begin{eqnarray}
R(x,t) =
 \int_{- \infty}^{+ \infty}
{d (\hbar \omega) \over 2\pi}
\frac{ g(\omega)}
{ \left( [ \hbar \omega -\lambda x]^2+\frac{[\hbar \Gamma(t)]^2}{4} \right)^2},
\end{eqnarray}
\begin{eqnarray}
R_1(x,t) =
 \int_{- \infty}^{+ \infty}
{d (\hbar \omega) \over 2\pi} g(\omega)
\frac{ (\hbar \omega -\lambda x)  }
{ \left( [ \hbar \omega -\lambda x]^2+\frac{[\hbar \Gamma(t)]^2}{4} \right)^2},
\end{eqnarray}
where $\hbar \ g(\omega)=-\partial_{\omega}f(\omega)$.\\
We note that while $N^{(0)}$ depends on the Fermi distribution $f(\omega)$, $N^{(1)}$ depends on its derivative.

One can calculate the adiabatic expansion for the current $J_{\alpha}(x,v,t)$ from the lead $\alpha$ to the dot.
We emphasize that the zeroth order expansion for the current vanishes (due to absence of voltage bias) and then
$J_{\alpha}(x,v,t) \simeq J_{\alpha}^{(1)}(x,v,t)$, where

\begin{equation}
J_{\alpha}(x,v,t) \simeq -e  \left( \lambda v [\hbar \Gamma_{\alpha}(t)] V(x,t) + [\hbar \dot{\Gamma}_\alpha(t)] V_1(x,t) \right),
\end{equation}
with $e$ the modulus of electron charge,
\begin{equation}
V(x,t) = \int_{- \infty}^{+ \infty}
{d (\hbar \omega) \over 2\pi}  \frac{ g(\omega)}
{ [ \hbar \omega -\lambda x]^2+\frac{[\hbar \Gamma(t)]^2}{4} },
\end{equation}
and
\begin{equation}
V_1(x,t) = \int_{- \infty}^{+ \infty}
{d (\hbar \omega) \over 2\pi}  \frac{ g(\omega) [ \hbar \omega - \lambda x]}
{ [ \hbar \omega -\lambda x]^2+\frac{[\hbar \Gamma(t)]^2}{4} }.
\end{equation}
Within the adiabatic expansion, the charge conservation is valid at the zeroth order of the dot occupation:
$e \dot{N}^{(0)}(x,t)=J_L(x,v,t)+J_R(x,v,t)$.

\subsection{Langevin equation for the oscillator}

In this subsection, we analyze the dynamics of the oscillator within the adiabatic regime. The effect of the electron bath
and the electron-oscillator coupling gives rise to a stochastic Langevin equation for the vibrational mode.
This equation is characterized by a position and a time dependent dissipation term, and a multiplicative noise.

Within the adiabatic limit, the force can be decomposed as:
\begin{equation}\label{for}
F(x,v,t)=F^{(0)}(x,t)+F^{(1)}(x,v,t).
\end{equation}
The zero order force $F^{(0)}(x,t)$ represents the "static" part sensitive to the average charge occupation
\begin{equation}
F^{(0)}(x,t)=-kx -\lambda N^{(0)}(x,t),
\end{equation}
with $N^{(0)}(x,t)$ given in Eq. (\ref{enne0}). The first order "dynamic" term $F^{(1)}(x,v,t)$ is sensitive to charge fluctuations, and it contains not only a dissipative term proportional to the velocity, but also a very complex non-linear term due to the effects of the pumping:
\begin{equation}
F^{(1)}(x,v,t)=-\lambda N^{(1)}(x,v,t)=-A(x,t)v + B(x,t) \dot{\Gamma}(t),
\end{equation}
with the damping coefficient $A(x,t)$ (positive definite) and $B(x,t)$ taken from
Eq.(\ref{enne1}):
\begin{equation}
A(x,t)= \frac{\hbar \lambda^2}{2}  [\hbar \Gamma(t)] R(x,t),
\end{equation}
and
\begin{equation}
B(x,t) = - \frac{\hbar^2 \lambda}{2} R_1(x,t).
\end{equation}
We point out that the pumping term introduces a complex forcing contribution dependent on the position $x$
through $B(x,t)$.

In the adiabatic limit, exploiting the effect of the \textit{fast} electronic environment on the oscillator motion,
one derives the following fluctuating term
\begin{equation}\label{fluforce}
\langle\delta\hat F(t)\delta\hat F(t')\rangle=
D(x,t) \delta(t-t'),
\end{equation}
where, due to the absence of electron voltage bias, $D(x,t)=2 K_B T A(x,t)$, that is the fluctuation-dissipation
condition is verified for each fixed position $x$ and time t.

The resulting Langevin equations for the oscillator dynamics becomes
\begin{eqnarray}\label{Langevin1}
\dot{x}&=&v, \\ \nonumber
m \dot{v} &=& - A(x,t) v+F_{tot}(x,t)+ \sqrt{D(x,t)}\xi(t), \\ \nonumber
\langle\xi(t)\rangle&=&0,\;\;\;\;\langle\xi(t)\xi(t')\rangle=\delta(t-t'), \nonumber
\end{eqnarray}
where $F_{tot}(x,t)$ is the deterministic part of the force
\begin{equation}
F_{tot}(x,t)=F^{(0)}(x,t)+B(x,t) \dot{\Gamma}(t),
\end{equation}
and $\xi(t)$ is a standard white noise term. In the limit of zero temperature
$D(x,t)=0$, so that we have a deterministic equation.

In the following, we will use these equations
to investigate the role of the deformable quantum dot on the pumped
current. Specifically, we will study how the quantum dot softness affects the
dependence of the pumped current on the temperature, pumping frequency, gate voltage, and the
phase difference between the pumping perturbations.

From the solution of the Langevin equation, one can calculate the oscillator distribution function $P(x,v,t)$, hence all
the properties of the oscillator. Moreover, one can determine the time behavior of an electronic observable $O_{el}(x,v,t)$:
\begin{equation}
\langle O_{el}\rangle(t)=\int \int d x d v P(x,v,t) O_{el}(x,v,t).
\end{equation}
If the pumping terms $u_{\alpha}(t)$ are periodic over $T_P$, then the coefficient $A(x,t)$ and the force $F_{tot}(x,t)$
exhibit the same periodic behavior. Consequently, we have found that the solutions of Langevin equation reproduce
themselves after one period, apart from a constant factor. In Appendix 1, we will discuss the numerical convergence of the results obtained 
from the solution of the Langevin equation (\ref{Langevin1}). Most numerical results reported in this work will be shown with symbols whose size 
will provide an estimate of the numerical error involved in the calculation.

In the following, we study the pumping with a very simple perturbation periodic over a period $T_P$ \cite{note-gamma}: $u_{\alpha}(t)=1+S \cos{(\omega_P t +\phi_\alpha)}$, with $\omega_P=2 \pi/T_P$ the pumping frequency and $\phi_\alpha$ the pumping phase of the lead $\alpha$. Therefore, we define $\Delta \phi = \phi_L - \phi_R$ as the phase shift between the two pumping parameters. We calculate the average $\bar{O}_{el}$ of a time-dependent electron quantity $\langle O_{el}\rangle(t)$, such as the dot occupation or the current, on a period $T_P$ as
\begin{equation}
\bar{O}_{el}=\frac{1}{T_P} \int_{0}^{T_P} d t \langle O_{el}\rangle(t).
\end{equation}

In the regime of adiabatic pumping, one has $\omega_P \ll\Gamma_0$, and $\omega_0 \ll \Gamma_0$, so that the dimensionless ratio
$ r_P= \omega_P / \omega_0 $ is of the order of unity. The regime of weak pumping is defined by the condition $S\ll1$.
Through the paper, we will assume $\omega_0=0.1 \Gamma_0$. We will measure lengths in units of $\lambda / k$,
times in units of $1/\Gamma_0$, and energies in units of $\hbar \Gamma_0$.

\section{Results}

\subsection{Distribution probabilities of the vibrational mode in the presence of pumping}

In order to understand the role of a deformable quantum dot on a Thouless pump, it is useful to analyze preliminarily the distribution probabilities in the configuration space of the vibrational (center of mass) mode. As elucitated in the previous section, these are calculated selfconsistently under generic pumping strength and arbitrary electron-oscillator coupling.
In Fig.\ref{pisto}  (upper panel) we plot the reduced position probability distribution as a function of $x$
for different times away from the mechanical resonance. For comparison, the equilibrium stationary distribution
for the harmonic oscillator is shown with dotted line. The effect of the pumping-assisted mechanical deformation causes a shift of $\langle x \rangle$ and a redefinition of the probability distribution variance $\sigma^2_x(t)$. However, we note that the bell-shaped character of the unperturbed distribution is maintained. Very close to the mechanical resonance, Fig.\ref{pisto}  (lower panel), a relevant temporal variation of the position probability distributions appears and a bimodal distribution is produced (black line) when $P(x,v,t)$ is integrated over time. The features described above can be understood by analyzing the oscillation amplitude $R_e$ of an effective forced Duffing oscillator (DO) \cite{memory_qd} taken as an oversimplified version of Eqs. (\ref{Langevin1}). For a DO $R_e \approx S \times\Pi (\omega_P-\omega_0) \times \mathcal{C}(\Delta \phi)$, where the function $\Pi (\omega_P-\omega_0)$ is peaked on resonance and it is related to the modulus of the response function of the vibrational mode, while $\mathcal{C}(\Delta \phi)$ contains  information related to the oscillator response to the phase shift $\Delta \phi$ of the driving signals. In the upper panel of Fig. \ref{pisto} the oscillation amplitude $R_e$ is lowered by the off-resonant value of the response function $\Pi$ despite the moderate value of the pumping strength $S=0.5$; on the other hand (lower panel), on resonance, the oscillation amplitude $R_e$ is strongly amplified despite the smaller value of $S$. The above arguments suggest that the effective pumping cycle seen by the electrons is completely dominated by the vibrational response to the driving signals. The correctness of the above arguments can be qualitatively validated by neglecting the non-linear part of the problem and working with a simple harmonic oscillator.

\begin{figure}
\centering
\includegraphics[width=8cm,height=9.0cm]{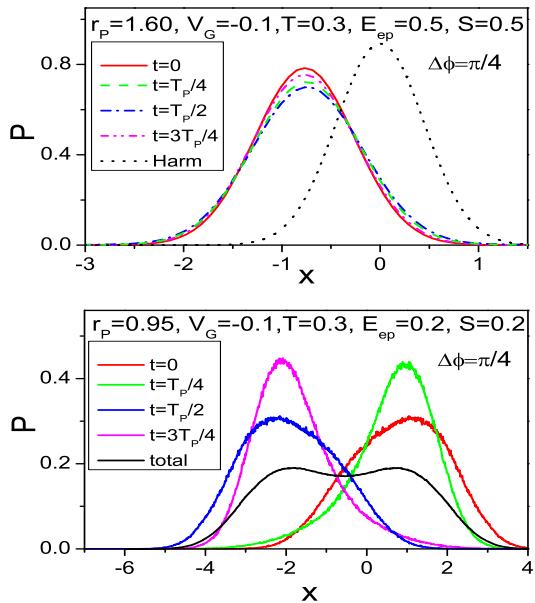}
\caption{(Color online) Upper Panel: Reduced position probability distribution as a function of $x$
for different times in the regime far from the mechanical resonance. For comparison, the equilibrium stationary distribution
for the harmonic oscillator is shown with dotted line.
Lower Panel: Reduced position probability distribution as a function of $x$
for different times in the regime very close to the mechanical resonance. The time average of the distribution is shown with black line and indicated
as "total".}\label{pisto}
\end{figure}

\subsection{Pumped charge}

In this Section we investigate the dependence of the magnitude of the leading contribution to the total charge pumped
per cycle $Q=T_{P} \langle{\bar J}_L\rangle$ on several model parameters. In particular, we will focus on the pumped charge dependence on temperature and driving frequency $\omega_{P}$. Moreover, we discuss the dependence of the charge pumped on the phase difference between the pumping perturbations $\Delta\phi$, the pumping strength $S$, the gate voltage $V_{G}$ and the electron-oscillator coupling $E_{ep}$ in order to optimize the amount of pumped charge
and to characterize possible non-linear effects.


As we will show in the following, the resonance condition is met for $\omega_{P}$  smaller than
$\omega_0$, since the dot occupation induces a strong softening of the bare frequency \cite{alberto2}. Therefore, the resonance takes place when the pumping frequency coincides with an effective oscillator
frequency $\omega_{eff}<\omega_0$. At the resonance, the distance between the maxima in the bimodal reduced position probability distribution (shown for example in the lower panel of Fig. \ref{pisto}) is the largest
and the pumped charge vanishes. \\
In the following, we will
denote $r_{eff}=\omega_{eff}/\omega_0$.
The renormalization of the bare resonator frequency is due to an effective potential of the oscillator and depends on the parameters of the system, among which the electron-oscillator coupling $E_{ep}$, the energy $V_G$, and the temperature $T$. In particular, at increasing temperature, the anharmonic contributions of the effective potential tend to become less important. This clearly induces a modification of the resonance frequency which approaches $\omega_0$ at high temperature.

In Fig. \ref{logvqplot} we plot the pumped charge $Q$ as a function of the pumping frequency $\omega_P$, hence as a function of $r_P$ ($r_P=\omega_P/\omega_0$) for different temperatures.
The $Q$ vs $r_P$ curves show a resonance peak close to $\omega_P \approx \omega_0$. The point where the pumped charge vanishes can be correctly assumed as the position
of the renormalized effective frequency $r_{eff}$. For example, one finds $r_{eff} \simeq 0.953$  at $T=0.3$}. A similar behavior is obtained by describing the vibrational mode in terms of a simple harmonic oscillator (see Fig. 4 of Ref.\cite{deformable_qd}).
In the following, we will focus our analysis to the region close to the resonance.
In the inset of Fig. \ref{logvqplot} the maximum   pumped charge (in logarithmic scale) vs the temperature is shown.
At increasing temperatures the pumped charge decreases even though it remains always higher than the value obtained in absence of the electron-oscillator coupling $E_{ep}$. While in the presence of the electron-oscillator coupling $E_{ep}=0.2$ an exponential decay with temperature is observed, for $E_{ep}=0$ the pumped charge decays more rapidly than a simple exponential\cite{note-charge}. The above result indicates that the presence of a vibrational mode provides an amplifying mechanism of the external driving signals assisting  the pumping mechanism. From the experimental point of view the slower decay of the pumped charge with temperature would favour the observation of the pumped current at temperatures higher than $mK$, overcoming the problem of measurements at cryogenic temperature. Finally, we point out that, on resonance, the pumped charge Q at finite $E_{ep}$ is amplified compared to the value at $E_{ep}=0$, while the situation is inverted out of resonance.

\begin{figure}
\centering
\includegraphics[width=8cm,height=9.0cm]{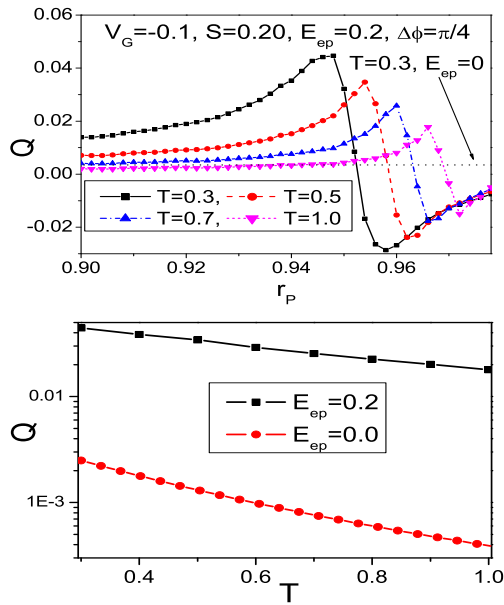}
\caption{(Color online) Upper Panel: The pumped charge Q as a function of the external frequency for different temperatures. For comparison,
the pumped charge for $E_{ep}=0$ is shown with dotted line. Lower Panel: The value of the charge at the maximum as a function of the temperature
is compared with the pumped charge for $E_{ep}=0$.}\label{logvqplot}
\end{figure}

In Fig.\ref{qvsphase} we plot the pumped charge Q as a function of the phase difference $\Delta \phi$ for different values of the external frequency, hence for different ratios $r_P$.
This plot is the analogue of the previous  figure at $T=0.3$.
We stress that close to the resonance $r_{eff} \simeq 0.953$, the amplitude of the pumped charge gets smaller (therefore Q tends to vanish for all $\Delta \phi$), while, crossing the resonance, a phase shift of $\pi$  is observed (see  Ref.\cite{deformable_qd} and Appendix 2).
Looking at Fig.4 one observes a $\sin (\Delta \phi)$ behavior only away from the resonance. Approaching the resonance condition, first one notes
slight deviations from the $\sin (\Delta \phi)$ due to dynamical effects and than a sudden sign change originated by a phase shift of $\pi$.
Away from the resonance, the behavior of the pumped charge over a broad range of frequency at finite temperature can be described by
\begin{eqnarray}\label{eq:q_osc}
Q &\propto& E_{ep}\Pi(\omega_P-\omega_{eff})\sin(\phi_D)\sin(\Delta \phi)
\end{eqnarray}
where $\Pi (\omega_P-\omega_{eff})$ is the modulus of the response function of the oscillator, while $\phi_D \sim \arctan(Q_{f}^{-1}\omega_P/(\omega_{eff}^2-\omega_P^2))$ is a dynamical phase shift which depends on  the effective quality factor $Q_{f}$ of the resonator. The phase shift of $\pi$, taking place on resonance, is fully explained by the functional dependence of the $\phi_D$ on $\omega_P$.
We also note that since we are in a symmetric case ($\Gamma_{L}=\Gamma_{R}$), the pumped charge is zero when $\Delta \phi=0$ and no rectification terms appear in the pumped charge. The analytical expression of the phase shift model given in  Eq. (\ref{eq:q_osc}) is inspired to the result given in Eq.(11) of Ref.\cite{deformable_qd} where the weak pumping regime and the zero temperature limit are assumed. In order to analytically validate Eq. (\ref{eq:q_osc}) in the presence of force fluctuations at finite temperature, the linear response theory for a stochastic system is required. This analysis, however, goes beyond the purposes of this work. It is worth mentioning that the sinusoidal form of the charge-phase relation expressed in Eq. (\ref{eq:q_osc}) is originated by the finite temperature effects which are able to weaken high harmonics contributions that are instead present in the zero temperature limit (see the Appendix 2 for details).


\begin{figure}
\centering
\includegraphics[width=9cm,height=7.0cm]{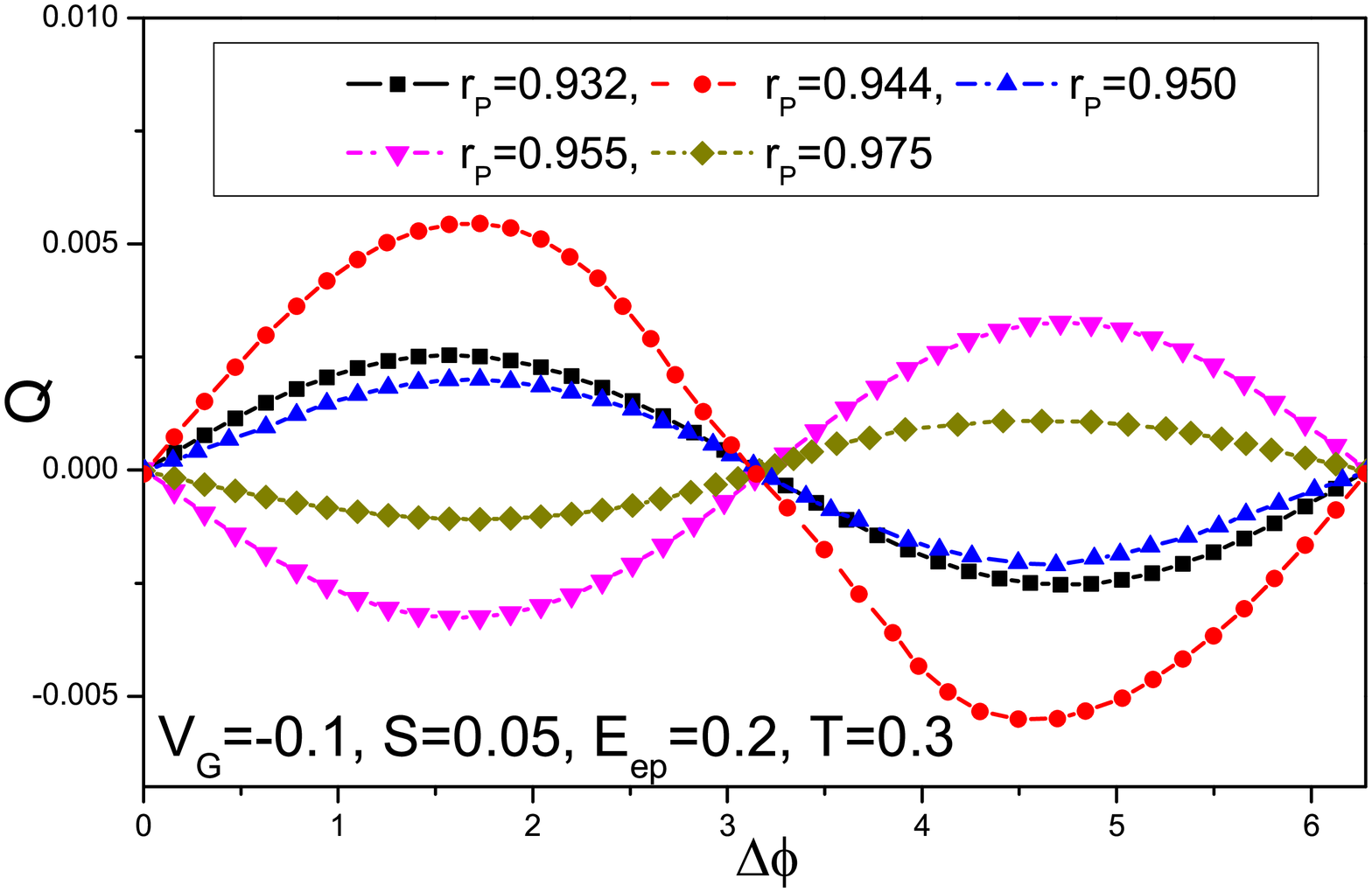}
\caption{(Color online) The pumped charge Q as a function of the phase difference $\Delta \phi$ for different values of the external frequency, hence for different ratios $r_P=\omega_P/\omega_0$.}
\label{qvsphase}
\end{figure}

While for weak pumping strength (the case $S=0.05$ has been shown in Fig. \ref{qvsphase}), a sinusoidal charge-phase relation is expected, further harmonics can develop  in the moderate pumping regime ($S>0.20$). In the latter condition, the oscillation amplitude of the vibrational mode increases and explores the non-linear region of the oscillator potential.
In Fig.\ref{q_weak_vs_strong}, where the $Q$ vs $\Delta \phi$ curves are computed in the moderate pumping regime ($S=0.5$), the deviation from the sine function becomes dramatic when approaching the resonance at $r_P=0.950$. As shown, the pumped charge increases in the strong coupling regime but its value does not reach a quantized value for the considered parameters range. The lack of quantization of the pumped charge per cycle can be ascribed to the difficulty in achieving a sufficient modulation amplitude of the tunneling rates compared to the static part. Indeed, apart from the time dependence induced by the vibrational mode, the pumping signals produce a life time modulation of the form $\Gamma_{\alpha}(t)\approx \Gamma^{0}_{\alpha}+\Gamma^{\omega}_{\alpha}\cos(\omega_P t+\phi_{\alpha})$ \cite{note-gamma}. Due to the form of $u_{\alpha}(t)$ the pumping ratio $\rho(S) \equiv \Gamma^{\omega}_{\alpha}/\Gamma^{0}_{\alpha}=\frac{4S}{1+S^2}$ is a bounded function of $S$, i.e. $0<\rho(S)\leq2$, and its value determines the weak or strong pumping regime. The strong pumping regime, defined by $\rho(S)>1$, is obtained in the interval $S \in[2-\sqrt{3},2+\sqrt{3}]\approx[0.268,3.732]$, while the maximum pumping ratio $\rho_M=2$ is reached for $S=1$. Thus the value of $\rho_M$ could be not sufficiently strong to reach the charge quantization condition.
Alternatively, the noise-induced transformation of the \textit{pumping cycle} into a complex trajectory contained within a \textit{circular crown} in the driving space $(\Gamma_L(t),\Gamma_R(t))$ can reduce the phase coherence required to maximize the transferred charge.

\begin{figure}
\centering
\includegraphics[width=9cm,height=7.0cm]{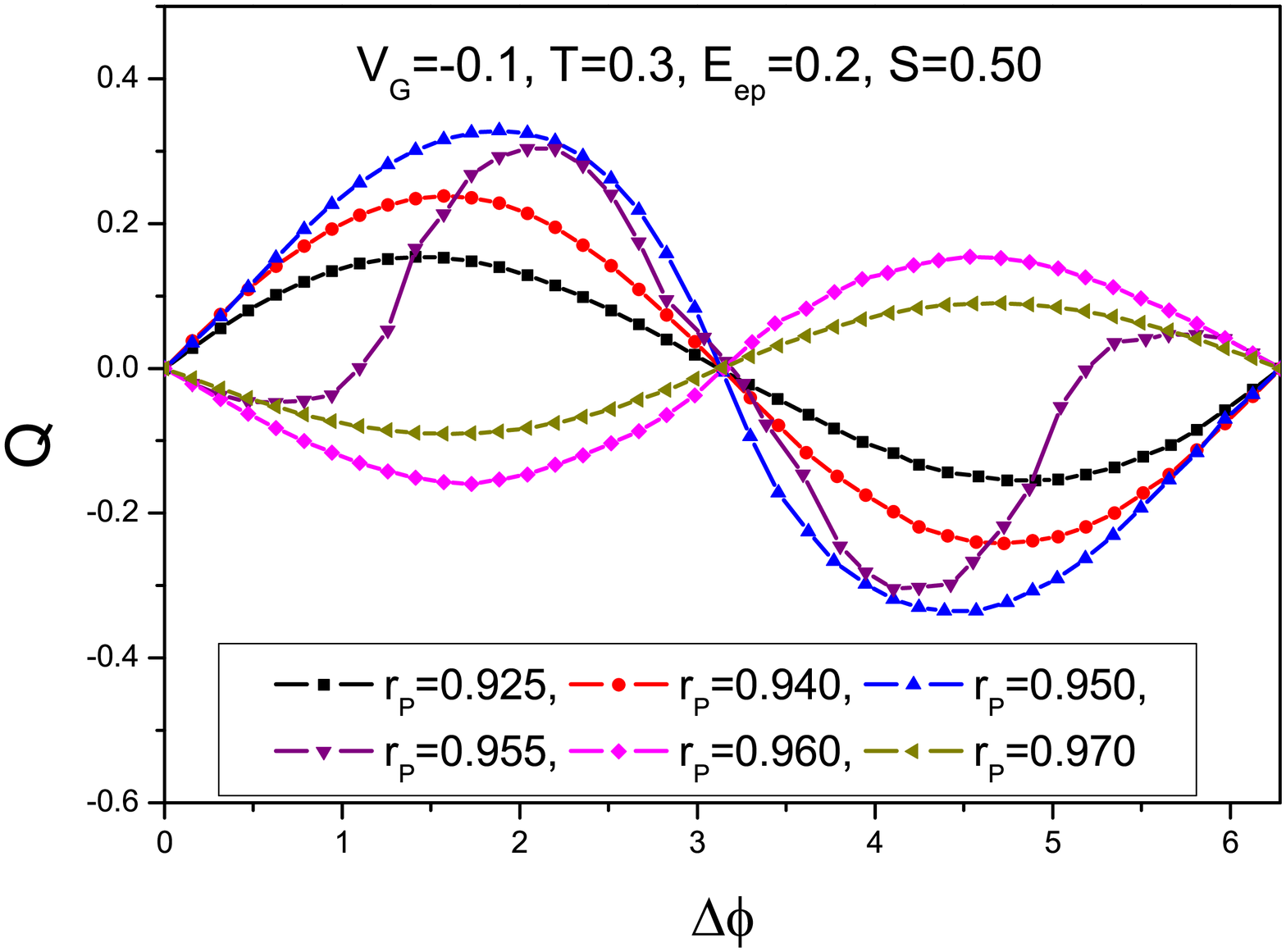}
\caption{(Color online) Upper Panel: The pumped charge Q as a function of the phase difference $\Delta \phi$ for different values of the pumping strength $S$. Lower Panel:  The pumped charge Q as a function of the phase difference $\Delta \phi$ for different values of $r_P$ for a larger value of pumping strength $S$.}\label{q_weak_vs_strong}
\end{figure}

In Fig. \ref{q_vs_ep} the pumped charge Q is plotted as a function of $V_G$ for different values of $r_P$. The charge behaves symmetrically for positive and negative values of $V_G$ and, for fixed $r_P$, a change of sign of $Q$ vs $V_G$ is observed.
Differently from the conventional Thouless pumping, we find an interesting behavior consisting of a rapid sign change of the pumped charge as a function of $V_G$. The distance of the non-trivial inversion points from $V_G-E_{ep}=0$ is an increasing function of $r_P$. In particular the gate values $V_G^{\ast}$ where the sign changes occur are determined by the resonance condition which is found by solving the equation  $r_P=r_{eff}(V_G^{\ast})$. The function $r_{eff}(V_G^{\ast})$ vs $V_G^{\ast}$ asymptotically ($|V_G^{\ast}| \rightarrow \infty$) saturates to the  bare resonance frequency of the oscillator, while it presents a single minimum close to $V_G^{\ast}-E_{ep}=0$.

In Fig. \ref{q_vs_ep} (upper panel), the values $r_P=0.92$ and $r_P=0.94$ are always smaller than  $r_{eff}$. The value $r_P=0.95$ satisfies the condition  $r_P=r_{eff}$ for $V_G-E_{ep} \simeq 0$,
while for $r_P=0.96$ and $r_P=0.97$, Q suddenly changes sign at finite values of $V_G-E_{ep}$. For larger values of $r_P$
(not shown in Fig. \ref{q_vs_ep}, but analogously to the lower panel of Fig. \ref{figa2} in Appendix 2), the pumped charge flattens, then,
for $r_P$ close to unity, it tends to small negative values.}
These results show that $r_{eff}$ approaches one for large $V_G$ while there is a lower limit for the renormalization of the resonance frequency, in agreement with Eq. (9) of Ref.\cite{deformable_qd}.
A detailed analysis of $r_{eff}$  as a function of $V_G$  at $T=0$ is reported in the Appendix 2.\\
In the lower panel of Fig.\ref{q_vs_ep} the pumped charge Q is plotted as a function of the electron-oscillator coupling  $E_{ep}$ for different values of the pumping strength $S$. Going from small to moderate values of the pumping strength $S$
an amplifying behavior at $E_p=0.2$ before the sign change is visible.


\begin{figure}
\centering
\includegraphics[width=8cm,height=9.0cm]{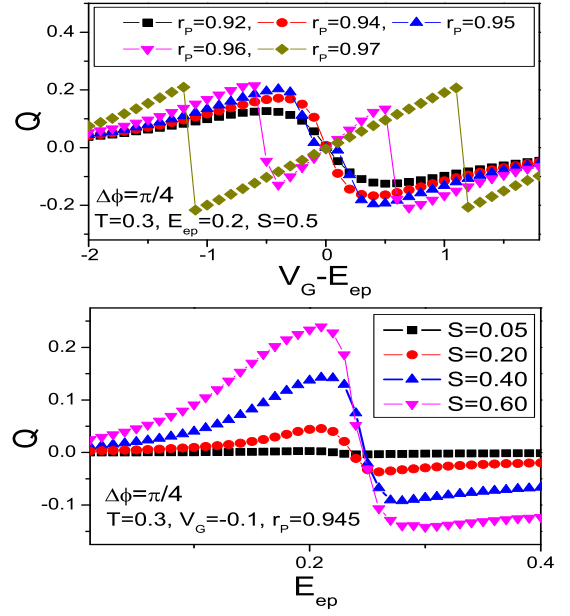}
\caption{(Color online) Upper Panel: The pumped charge Q as a function of $V_G-E_{ep}$ for different values of $r_P$. Lower Panel:  The pumped charge Q as a function of the electron-oscillator coupling  $E_{ep}$ for different values of the pumping strength $S$. Notice that the pumping strength $S=0.6$ corresponds to the driving ratio $\Gamma^{\omega}_{\alpha}/\Gamma^{0}_{\alpha}\approx1.765$, which is very close to the maximal value of 2.}\label{q_vs_ep}
\end{figure}

It is also clear that the variation of $E_{ep}$ renormalizes the bare mechanical resonance frequency of the system thus providing the opportunity to go from a off-resonance to a resonant condition at fixed pumping frequency. Moreover, the force fluctuations induced by the charging effects of the molecular energy level do not destroy the quantum pumping mechanism.


\section{Conclusions and Discussions}
We have employed a non-equilibrium Green's function approach to study the quantum pumping through a molecular level coupled to a classical vibrational mode governed by a stochastic dynamics. We have shown  that the presence of dissipation and noise doesn't destroy the quantum pumping mechanism and, even, reinforces it when the pumping frequency is tuned close to the resonance. Actually, a wide parameters range has been identified where the vibrational mode provides an amplifying mechanism that cooperatively assists the quantum pumping and the simulation results can be often understood in term of
an effective frequency which renormalizes the bare resonator frequency $\omega_0$. When looking at the pumped charge as a function of $\Delta \phi$ a non-conventional behavior
is found compared to the Thouless pump. The behavior is clearly different from a $\sin (\Delta \phi)$ and the contributions of more harmonics are already clear
at $T=0$ (see Appendix 2) as a results of the non-linear dynamics.
The robustness of the noise-assisted quantum pumping has also been investigated as a function of the temperature and an exponential decay of the pumped charge in the presence of the electron-oscillator coupling has been found.

Our results clarify the effect of a classical perturbation (a single resonant harmonic oscillator) on the quantization of the charge pumped in a cycle by a Thouless pump. We notice that our study is complementary to that reported in Ref. \onlinecite{silva-fioretto2008} where the coupling of a  bosonic bath to a Thouless pump has been studied in the low temperature regime in a fully quantum mechanical approach. We emphasize that in our case the temperature is larger than the oscillator frequency, while in Ref.\onlinecite{silva-fioretto2008} the typical cut-off frequency of the bosonic bath is much larger than temperature. The difference between the two approaches comes out clearly in the temperature dependence of the pumped charge. In the case of a quantum bosonic bath,  a power law decay of the pumped charge from the quantized value is observed, while, in the case of a single classic resonator, as mentioned, we find a simple exponential decay.  From this point of view, the differences can be interpreted as the result of the quantum nature of the bosonic bath against the classical nature of the resonator present in our set-up. In this framework we can look at our system from a different point of view. It  can be considered as a quantum system (the Thouless pump) coupled to the simplest classical measurement system (the oscillator). The oscillator realizes a continuous measurement process and can be seen as the sensitive element of the macroscopic physical system that measures the charge density inside the pump. The effect of the coupling of the quantum and classical system is twofold. From one side, the continuous measurement process induces a moderate amount of decoherence in the quantum system and
the oscillator acts as an \textit{engineered environment}.
 On the other hand, the quantum system interacts with the classical degree of freedom of the measurer affecting its
dynamics with stochastic terms coming from the quantum correlations.  The above findings are relevant in understanding the
quantum dynamics close to the boundary with the classical realm.  In particular, the interaction Hamiltonian
$H_{int}=\lambda \hat{n}\hat{ x}$ can be tuned via the coupling constant $\lambda$ whose value is usually related
to the geometry of the device (effective capacitance) or is affected by gate voltages.
In the latter case, the appropriate tuning of the parameter $\lambda$ (and, consequently, $E_{ep}$) allows to introduce in a
controlled way a definite amount of decoherence with important implications on the optimization of a quantum state.
Compared to \textit{engineered environments} obtained by using electronic or bosonic degrees of freedom (bath),
our proposal guarantees the adiabatic nature of the detector dynamics. The proposed scheme describes a dynamical process
of measurement of a quantum system, beyond the wavefunction collapse picture.

\appendix*

\begin{appendix}

\section{Appendix 1: Numerical Procedure}

In this Appendix we will discuss the numerical procedure followed in this work. We will focus on the numerical convergence of the physical quantities, in particular the pumped charge Q. 

All the results discussed in the main text derive from the numerical solution of the Langevin equation (\ref{Langevin1}). We solve this second-order stochastic differential equation extending a fourth-order stochastic Runge-Kutta algorithm \cite{Honey1,Honey2}. In order to solve the second-order equation with multiplicative white noise, we decompose the problem into a set of three first-order differential equations \cite{Bao}. The third equation takes into account the effect of spatial dependence of the noise, involving a non-multiplicative noise term. For our
simulations we have fixed a time step $t_s = 0.1/ \omega_0$ and set long simulation times up to  $10^9 t_s$. Within these
settings, the algorithm shows an excellent stability and reaches the convergence for all the oscillator and electronic quantities. In order to average upon statistically independent events, we have sampled the values of x(t) and v(t) every 10 time steps.

In the Langevin equation (\ref{Langevin1}), the fluctuating term $D(x,t)$ gets smaller with decreasing the temperature $T$ and the electron-oscillator coupling $E_{ep}$. Therefore, the parameter regime with very small values of $T$ ($T<0.1$) and $E_{ep}$ ($E_{ep}<0.1$) is not easily accessible since it requires longer and longer dynamics. This regime has been discussed in Appendix 2 where, at zero temperature, in the absence of the fluctuating term, the features of the model can be obtained for very weak electron-oscillator coupling $E_{ep}$.
The central quantity discussed in the main text, the pumped charge Q, is affected by the mechanical resonance (where it goes to zero) and by the temperature increase (where it strongly decreases). Indeed, accurate numerical convergence is needed for the estimation of the pumped charge particularly in the regime of weak pumping. For this reason, in this Appendix, we discuss the features of the pumped charge within the numerical procedure.

\begin{figure}
\centering
\includegraphics[width=8cm,height=9.0cm]{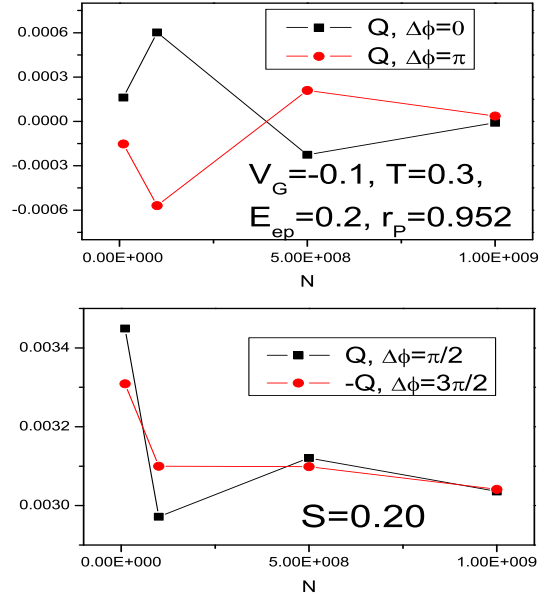}
\caption{(Color online) Upper Panel : The pumped charge Q as a function of the number N of time steps at $\Delta \phi=0, \pi$. Lower Panel : The pumped charge Q as a function of the number N of time steps at $\Delta \phi=\pi/2, 3 \pi/2$.}\label{figap1}
\end{figure}

In Fig. \ref{figap1}, we report the convergence of Q with increasing the number of time steps in a parameter regime close to the mechanical resonance. In particular, we show the pumped charge Q at values of the phase difference $\Delta \phi$ where the symmetry of the problem constrains the results. In the upper panel of Fig. \ref{figap1}, we consider the cases $\Delta \phi=0$ and $\Delta \phi=\pi$, where, by symmetry considerations, one expects the vanishing of Q. We point out that the convergence is not fast. Actually, for $N=5 \times 10^8 $ time steps, the charges at
$\Delta \phi=0$ and $\Delta \phi=\pi$ still differ in terms of $5 \times 10^{-4}$. Only for $N=10^9 $ time steps, the two charges are equal within an error of  $4 \times 10^{-5}$, which is one order of magnitude smaller than the case $N=5 \times 10^8 $. 

The values of the charges shown in the upper panel of Fig. \ref{figap1} tends to be always opposite. This is due to the relation of the charge at $\Delta \phi$ and $2 \pi - \Delta \phi$. Actually, these relations are also found for finite values of the charge. In the lower panel of Fig. \ref{figap1}, we report Q at $\Delta \phi=\pi/2$ and -Q at $\Delta \phi=3 \pi/2$. They tend to become equal with increasing N. We point out that the convergence for these values of $\Delta \phi$ improves quite sensitively. For $N=5 \times 10^8 $ time steps, the charges are different only in terms of $2 \times 10^{-5}$. Moreover, for $N=10^9 $ time steps, the error in the estimate  of the charges is smaller than $10^{-5}$. From the comparison of the charges at several values of $\Delta \phi$, we have checked that the error at  $\Delta \phi=0$ and $\Delta \phi=\pi$  is the largest, therefore it sets an estimate of the numerical resolution within our procedure. In the main text, most results have been plotted with symbols whose size provides an estimate of the error. When the charge as a function of the phase difference $\Delta \phi$ has been plotted, we have used the estimate of the error at $\Delta \phi=0$.

\end{appendix}

\begin{appendix}
\section{Appendix 2: Results at T=0}

The result of this Appendix helps clarifying some points explained in the main text. In particular, we analyze the system at $T=0$ focusing on the parameters region close to the mechanical resonance. In this case, the Langevin equation reduces to a deterministic equation.  We solve this second-order differential equation with a fourth-order Runge-Kutta algorithm. In analogy with the numerical procedure at finite temperature, we have fixed a time step $t_s = 0.1/ \omega_0$ and considered long simulation times up to $10^9 t_s$. The numerical accuracy of the results is very good. Symbols are used in the plots to indicate an estimate of the error within the numerical procedure.

The analysis at $T=0$ allows to describe in more detail the response close to the mechanical resonance in the limit when the electron-oscillator coupling $E_{ep}$ is small, even though some quantities, like the effective oscillator frequency $r_{eff}$, show a temperature dependent renormalization not captured in this limit. Despite these limitations, the $T=0$ case treated here clarifies the system behavior.

\begin{figure}
\centering
\includegraphics[width=9cm,height=7cm]{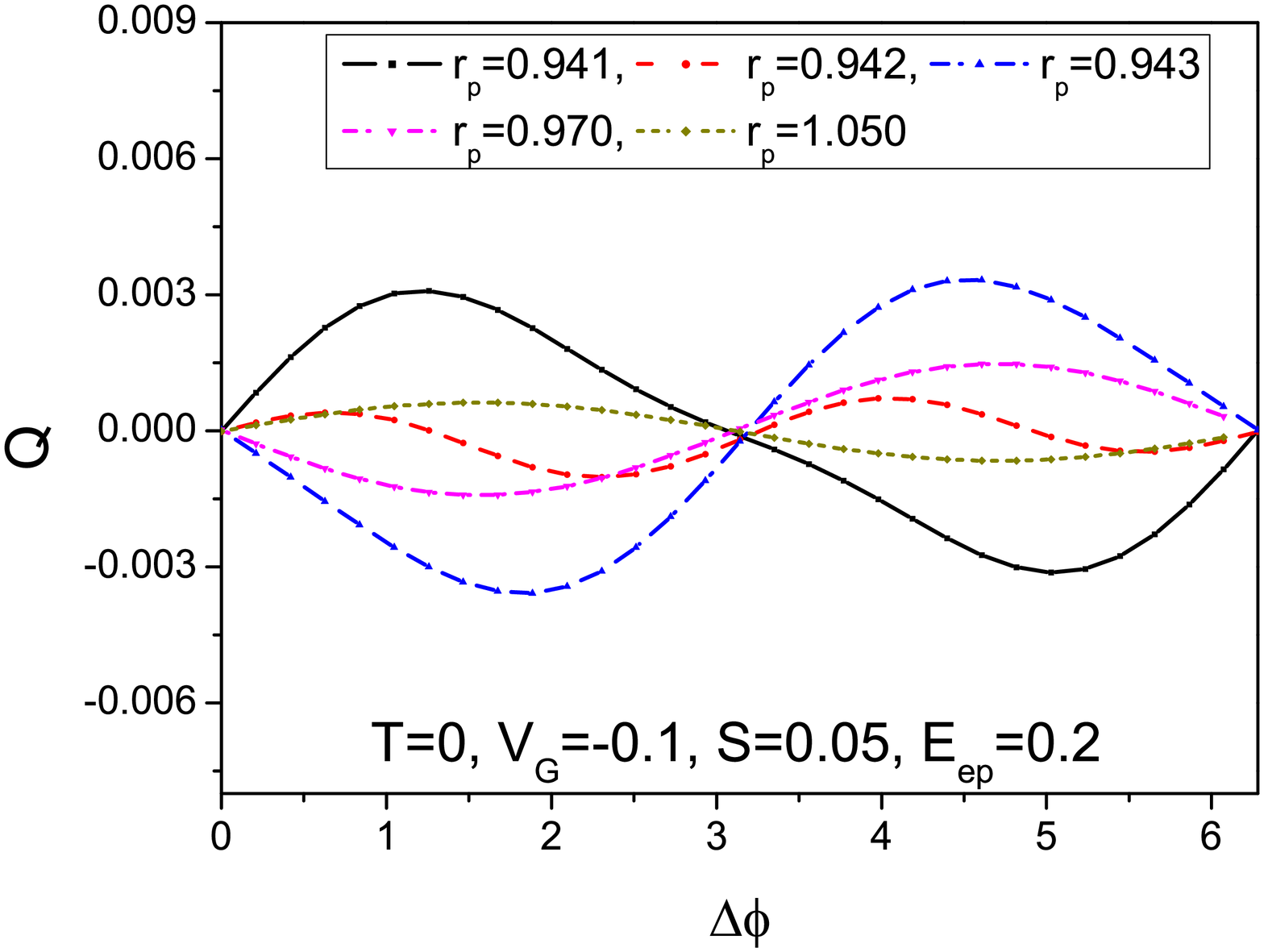}
\caption{(Color online) The pumped charge Q as a function of the phase difference $\Delta \phi$ for different values of $r_P=\omega_P/\omega_0$ at $T=0$.}\label{figa2}
\end{figure}

In comparison with the results obtained at $T=0.3$ and reported in Fig. \ref{qvsphase} (with $r_{eff}$ close to $0.952$), in Fig. \ref{figa2} we focus on the resonance at $T=0$ maintaining the remaining parameters as fixed in Fig.\ref{qvsphase}. Fig. \ref{figa2} ( $T=0$ case) clearly shows that $r_{eff} \simeq 0.942$, while, crossing the resonance, a phase shift of $\pi$ is observed. Moreover, even if the amplitude of the pumped charge is small, we recognizes additional nodes in the $Q$ vs $\Delta \phi$ curve, evidencing the contribution of a second harmonic term. We have checked that higher harmonics are present in the time evolution of the quantity $\langle Q\rangle(t)$. Therefore, at the resonance, where the amplitude of the oscillator is large, the pumped charge is strongly affected by anharmonic effects. As soon as one goes away from the resonance, the nonlinear contributions quickly get depressed.
Fig. \ref{figa2} also clarifies the behavior of the pump far from the resonance. While the behavior at $r_P=0.970$ is similar to that just after the resonance (but more sinusoidal), at $r_P=1.050$ there is a phase change of the sinusoidal curve. Indeed, with increasing $r_P$, an additional phase shift close to $r_P \simeq 1$ is detected and, consequently,  the system recovers the behavior preceding the resonance.

\begin{figure}
\centering
\includegraphics[width=8cm,height=9.0cm]{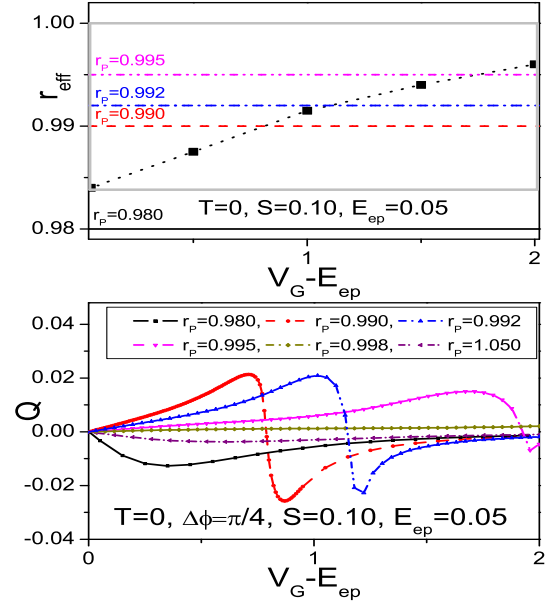}
\caption{(Color online) Upper Panel: The effective oscillator frequency $r_{eff}=\omega_{eff}/\omega_0$ as a function of $V_G-E_{ep}$ for $T=0$. The horizontal lines correspond to the values of $r_P$ considered in the lower panel. Lower Panel:  The pumped charge Q as a function of $V_G-E_{ep}$ for different values of $r_P$ at $T=0$.}\label{figa3}
\end{figure}

In the second part of the Appendix, we address the behavior of the pumped charge as a function of $V_G$.
In the upper panel of Fig. \ref{figa3}, we show the behavior of $r_{eff}$ as a function of $V_G$. The largest renormalization of the oscillator frequency occurs at $V_G-E_{ep}=0$ (in the case reported in figure $r_{eff} \simeq 0.984$). Moreover, $r_{eff}$ tends to one (i.e. the bare resonance frequency) for large $V_G$. The grey rectangle in the upper panel of Fig. \ref{figa3} delimits the region where the oscillator frequency is renormalized. In the same figure, the horizontal lines correspond to the values of $r_P$ considered in the lower panel. The crossing points between the pumping frequency $r_P$ and $r_{eff}$ selects the point $V_G-E_{ep}$ where the resonance takes place, i.e. where the pumped charge vanishes and changes sign. As reported in the main text, at the mechanical resonance $r_{eff}$, the phase shift of $\pi$ is well resolved as a function of $V_G-E_{ep}$.
In the lower panel of Fig. \ref{figa3}, we show that $r_P=0.980$ is smaller than the values in the grey area, therefore the pumped charge never approaches the resonance. However, $r_P=0.990$ is well within the grey area, therefore, the resonance has already occurred and the pumped charge has a different sign. For $V_G-E_{ep}$ larger than $0.8$, the system cannot approach the resonance, therefore the pumped charge has again a negative sign. For $V_G-E_{ep}$ around $0.8$, the crossing of the resonance takes place with a rapid change of the pumped charge.
Another point shown in the lower panel of Fig. \ref{figa3} is the behavior far from the resonance. With increasing $r_P$, the pumped charge flattens ($r_P=0.995$), then, for $r_P$ close to unity ($r_P=0.998$), it reaches small negative values. Finally, for $r_P$ larger than one   ($r_P=1.050$), the pumped charge recovers a positive sign and the small values characteristic of the regime before the resonance.

\end{appendix}





\begin{thebibliography}{99}
\bibitem{note-charge} The fitting curve of $Q$ vs $T$ for $E_{ep}=0$ is of the form $Q=\exp(a+b T+cT^2)$  with  $a=-2.21602$, $b=-1.46152$, e $c=0.27523$ for the values of the parameters considered in Fig.2.
\bibitem{note-gamma} Notice that the pumping driving $u_{\alpha}(t)$ originates an hybridization energy $\Gamma(t)=\Gamma_0 \{(1+S^2)+\sum_{\alpha}[2S\cos(\omega_P t+\phi_{\alpha})+\frac{S^2}{2}\cos(2\omega_P t+2\phi_{\alpha})]\}$ which contains a spurious second harmonic term. The latter contribution determines a violation of the monocromatic driving assumption usually made in the orthodox pumping theory which can originate important deviations for $S \gg 1$. Further harmonics are also generated by the non-linear dependence of the dot occupation $N(x,v,t)$ on $\Gamma(t)$.
\end{thebibliography}
\end{document}